\documentclass[aps,pre,showpacs,twocolumn,superscriptaddress]{revtex4-1}
\usepackage{amsmath}
\usepackage{amssymb}
\usepackage{graphicx}
  \usepackage{hyperref}
\usepackage[arrow, matrix, curve]{xy}
\usepackage{bm}
\usepackage{yhmath}
\usepackage{color}
\newcommand\diff{\mathrm{d}}

\usepackage[usenames,dvipsnames]{xcolor}
\hypersetup{colorlinks=true, linkcolor=BrickRed, urlcolor=blue!50!black, citecolor=blue!50!black}


\usepackage[normalem]{ulem}

\makeatletter
\newcommand\hide@visible[1]{%
  \bgroup\fboxsep=.3ex\colorbox{Gray}{begin hide}%
  #1\colorbox{Gray}{end hide}\egroup%
}
\newcommand\hide@hidden[1]{%
  \bgroup\fboxsep=.3ex\colorbox{Gray}{hidden text}%
}
\newcommand\hide@invisible[1]{}
\newcommand\makevisible{\let\hide\hide@visible}
\newcommand\makehidden{\let\hide\hide@hidden}
\newcommand\makeinvisible{\let\hide\hide@invisible}
\makeatother
\makehidden

\begin{document}


\title {Structural quantities of quasi-two-dimensional fluids}



\author{Simon Lang}
\author{Thomas Franosch}
\affiliation{Institut f\"ur Theoretische Physik, Leopold-Franzens-Universit\"at Innsbruck, Technikerstra{\ss}e~25/2, A-6020 Innsbruck, Austria}
\affiliation{Institut f\"ur Theoretische Physik, Friedrich-Alexander Universit{\"a}t Erlangen-N\"urnberg, Staudtstra{\ss}e~7, 91058 Erlangen, Germany}

\author{Rolf Schilling}
\affiliation{Institut f\"ur Physik, Johannes Gutenberg-Universit\"at Mainz,
 Staudinger Weg 7, 55099 Mainz, Germany}





\date{\today}
\begin{abstract}
Quasi-two-dimensional fluids can be generated by confining a fluid between two parallel walls with narrow separation. Such fluids exhibit an inhomogeneous structure perpendicular to the walls due to the loss of translational symmetry. Taking the transversal degrees of freedom as a perturbation to an appropriate 2D reference fluid we provide a systematic expansion of  the $m$-particle density for arbitrary $m$. To leading order  in the slit width this density factorizes into the densities of the transversal and lateral degrees of freedom.  Explicit expressions for the next-to-leading  order terms are elaborated analytically quantifying the onset of inhomogeneity.   
The case $m=1$ yields the density profile with a curvature given by an integral over the pair-distribution function of the corresponding 2D reference fluid, which reduces to its 2D contact value in the case of pure excluded-volume interactions. Interestingly, we find that the 2D limit is subtle and requires stringent conditions on the fluid-wall interactions. We quantify the rapidity of convergence for various structural quantities to their 2D counterparts.
\end{abstract}






\maketitle

\section{Introduction}

Confining fluids on length scales comparable to the typical interaction range introduces a competition between local particle near ordering and the layering induced by the particle-wall interaction eventually leading to a dimensional reduction. Of particular interest is the  influence of confinement on the structure of fluids~\cite{Alba:2006,Dietrich:1995,Lowen:2001,Lowen:2009,Evans:1990,Nygard:2012,Oguz:2012,Zwanikken:2013,Nygard:2013}, their phase equilibria, equilibrium phase transitions  and their critical behavior (see e.g. Refs.~\cite{Binder:1974,Fisher:1981,Vink:2006,Evans:1990,Liu:2010} and references therein) as well as their dynamic properties~\cite{Klafter:Restricted,Mittal:2008,Nugent:2007,Lang:2010,Gallo:2000a,Krishnan:2012,Watanabe:2011}.
The simplest realization of confinement consists of enclosing the fluid in a slit geometry between two parallel walls. For strong confinement only a few layers of particles fit between the plates. 
 The regime of strong confinement has been investigated by computer simulations   (see Ref.~\cite{Antonchenko:1984,Alejandre:1996,Schmidt:1996,Schmidt:1997,Fortini:2006,Gribova:2011,Qi:2013} and references therein)  and theoretically by employing suitable closures for the integral equation approaches~\cite{Antonchenko:1984,Henderson:1986,Adams:1989,Alejandre:1996,Henderson:1997,Schmidt:1997,Xu:2008a,Xu:2008b} or within density functional theory~\cite{Tarazona:1987, Rosenfeld:1996, Rosenfeld:1997, Goetzelmann:1997}.

In colloidal suspensions a range of plate separations can be investigated experimentally   by imposing a small inclination of one of the plates~\cite{Nugent:2007,Desmond:2009,Neser:1997,Spannuth:2012,Reinmueller:2013}. Alternatively the confinement may be achieved by bringing a large glass sphere in close contact with a planar wall and monitor the colloids confined within the gap~\cite{Eral:2009}. Structural information on the particle arrangements in confinement has also been collected using small angle X-ray scattering~\cite{Nygard:2009,Nygard:2012}, which experimentally allows to probe structural quantities even in the regime of quasi-two-dimensional fluids~\cite{Nygard:2009,Bunk:2007}.

Recently, it has been shown that in the limit of extreme confinement~\cite{Franosch:2012} a small parameter $n_0 L^2$ emerges, where $n_0$ is the number of particles per area and $L$ the  separation
of the walls. In particular, to leading order the transversal and lateral degrees of freedom (d.o.f.)  decouple
 which allows relating equilibrium phase transition lines with respect to the corresponding 2D reference system~\cite{Franosch:2012}. The next-to-leading order of thermodynamic quantities can be elaborated relying on a systematic cluster-expansion. Furthermore, an effective 2D two-body potential is obtained by integrating out the transversal degrees of freedom.

The goal of this work is to elaborate the influence of the transversal d.o.f. on structural quantities such as the density profile and the pair-distribution function in the regime of quasi-two-dimensional liquids. Thereby we will elucidate the role of the fluid-fluid and particularly of the fluid-wall interaction.  Despite the decoupling property~\cite{Franosch:2012} it will become obvious below that the 2D limit  can be rather subtle, depending on both, the physical quantity of interest and the fluid-wall interactions. We will demonstrate that for $L\to 0$ the coupling between the lateral and transversal d.o.f. can be treated as a perturbation which allows expanding the structural entities with respect to their transversal variables.  Here, we rely on the recently developed systematic expansion valid for $L\to 0$ ~\cite{Franosch:2012}. The leading-order term is the corresponding quantity of the homogeneous 2D reference fluid. Particularly, the analytically accessible corrections with respect to the 2D fluid  will be determined. These corrections describe the onset of inhomogeneity for confined fluids emerging from the homogeneous limit of a 2D fluid. The density and the density-density correlation function as probability densities permit to calculate the average of any one-particle and two-particle local observable in the limit of extremely confined liquids.

Our paper is organized as follows. In Sec.~\ref{Sec:Basic} the model is described and various structural quantities are introduced for which
the 2D  limit will be studied. This limit  will be discussed in Sec.~\ref{Sec:2DLimit}.  A summary and conclusions are given in the final section.
More technical calculations have been transferred to Appendices~\ref{Sec:acum}-\ref{Sec:ccluster}.

\section{Confined fluids: Basic structural quantities}\label{Sec:Basic}

We consider a liquid of $N$ structureless particles between two parallel, planar walls with cross section $A$ and separation $H$. We choose a coordinate system such that the $z$-direction is perpendicular to the walls located at $z= \pm H/2$. In $x$-$y$-direction periodic boundary conditions are assumed.
Then a point in configurational space  is specified by the set of coordinates $\vec{x}$ parallel and perpendicular to the wall  $\vec{x}_i = ( \vec{r}_i, z_i), i=1,\ldots N$. The walls are assumed to be impenetrable and the liquid-wall interaction to be additive
\begin{align}\label{eq:potential}
U(\{z\};L)  &=\sum_{i=1}^{N} {\cal U}(z_{i};L),
\end{align}
with the single-particle-wall interaction
\begin{align}\label{eq:potential_wall}
{\cal U}(z;L) &=
\begin{cases}
    \infty & \text{for} |z|> L/2,  \\
  {\cal U}_{-}(L/2+z)+{\cal U}_{+}(L/2-z) & \text{for} |z| \leq L/2.
\end{cases}
\end{align}
Here, we have introduced the effective wall distance $L$ as the transverse length accessible to the particles, and therefore distinguish between point particles and hard spheres
\begin{equation}\label{eq:L}
L=
\begin{cases}
    H-\sigma ,& \text{hard spheres},  \\
    H, & \text{point particles}.
\end{cases}
\end{equation}
The potential ${\cal U}_{+}(\mathcal{U}_{-})$ is a smooth interaction energy of a fluid particle with the right (left) wall, with possible singularities for vanishing argument.
With ${\cal U}_{-}(z)\neq {\cal U}_{+}(z)$ we further allow for asymmetric wall conditions.
 The mutual interactions between the fluid particles will be restricted to pair interactions ${\cal V}(\vec{x})\equiv {\cal V}(\vec{r},z)$, only depending on the magnitude  $|\vec{x}|$ of their relative distance
\begin{equation}\label{eq:pairpotential}
 V_0(\{ \vec{x} \})= \sum_{i < j}^{N} \mathcal{V}(\vec{x}_{i}-\vec{x}_{j}).
\end{equation}
 Then the total interaction energy reads
\begin{equation}
 V(\{ \vec{x}\};L)= V_{0}(\{\vec{x}\})+U(\{z\};L).
\end{equation}
In the following we investigate thermal averages using the configurational part of the canonical ensemble
 $\rho(\{\vec{x} \};L)=\exp[-\beta V(\{\vec{x} \};L)]/Z$, where  $Z(T,A,N,L) = \int [\prod_{i=1}^N \diff^2 r_i \diff z_i] \exp [-\beta V (\{\vec{x} \};L)]$ denotes the configurational partition function. Integrals over lateral coordinates are performed over the cross section $A$ and the perpendicular coordinates are confined to $-L/2 \leq z_i \leq L/2, i=1,\ldots,N$.  
Therefore the thermodynamic relevant volume is the accessible
volume $A L$ and the corresponding number density is given by $n=N/AL$.
Canonical averages are indicated by angular brackets $\langle (\cdots) \rangle$. The thermodynamic limit (TD-limit) $N \to \infty$, $A \to \infty$
is taken such that the area density $n_0=N/A$ remains fixed. Keeping the particle  density $n=N/AL$ constant implies that $n_0$
converges to zero for $L\to 0$, i.e. the 2D system becomes an ideal gas.  
  Therefore,  to ensure that correlations between the particles persist in the 2D limit one has to fix $n_0$ instead of $n$.
The dependence on the variables $T,A,N$ will mostly be suppressed, while we often highlight the dependence on the effective wall separation $L$.

A basic structural  entity characterizing the distribution of particles in a fluid is the $m$-particle density $\rho^{(m)}(\vec{r}_1z_1,...,\vec{r}_m z_m;L)$~\cite{Hansen:Theory_of_Simple_Liquids}. Within the confined geometry as introduced above, the $m$-particle density can be written in the subsequent form
\begin{align}\label{eq:mpdensity}
 \qquad \rho&^{(m)}(\vec{r}_1z_1,...,\vec{r}_m z_m;L)\nonumber \\
 & =\rho^{(m)}_{\perp}(z_1,...,z_m;L) \tilde\rho^{(m)}(\vec{r}_1z_1,...,\vec{r}_m z_m;L),
\end{align}
where we have introduced the $m$-particle density of the transversal degrees of freedom
\begin{eqnarray}\label{eq:mpdensity-a}
\rho^{(m)}_{\perp}(z_1,...,z_m;L) &=& \prod_{i=1}^{m}\rho^{(1)}_{\perp}(z_i;L), \nonumber\\
\rho^{(1)}_{\perp}(z_i;L)&=&  \exp[-\beta {\cal U}(z_{i};L)]/ z_{\perp}(L), \nonumber\\
z_{\perp}(L)&=&\int \diff z \exp[-\beta {\cal U}(z;L)],
\end{eqnarray}
and a reduced  $m$-particle density
\begin{align}\label{eq:mpdensity-b}
&\tilde\rho^{(m)}(\vec{r}_1z_1,...,\vec{r}_m z_m;L)= {\cal N}(L) \frac{N!}{(N-m)!} \nonumber\\
&\times \int  \left[\prod_{j=m+1}^N d^2r_{j}dz_{j}\rho^{(1)}_{\perp}(z_j;L)\right] \exp [-\beta V_0 (\{\vec{r}z \};L)] / Z_{\parallel}.
\end{align}
The normalization factor is given by
\begin{equation}\label{eq:normfactor}
{\cal N}(L) = Z_{\parallel}Z_{\perp}(L) /Z(L),
\end{equation}
with partial partition functions $Z_{\parallel} \equiv Z_{\parallel}(T,A,N)$ corresponding to the 2D reference fluid with pair potential
$V^{\parallel}_0(\vec{r}_1,...,\vec{r}_m) \equiv V_0({ \{\vec{r}\},0 })$ and $Z_{\perp} \equiv Z_{\perp}(T,N,L)= (z_{\perp}(L))^N$ to the ideal gas of the transversal d.o.f. with wall potential $U(\{z\};L)$ from Eq.~\eqref{eq:potential}. Note that Eq.~\eqref{eq:mpdensity} factorizes the trivial $z_i$- and $L$-dependence of
$\rho^{(m)}_{\perp}$ from the nontrivial one of $\tilde\rho^{(m)}$.
For $m=1$ one obtains the density profile
\begin{equation}\label{eq:profile}
n(z;L)= \rho^{(1)}(\vec{r}z;L),
\end{equation}
which is independent of $\vec{r}$ due to translational invariance along the walls. The pair-distribution function
$g(\vec{r}-\vec{r}',z,z';L)$ is related to $\rho^{(2)}(\vec{r}z,\vec{r}'z';L)$ by
\begin{align}\label{eq:pairdf}
g(\vec{r}&-\vec{r}',z,z';L) \negthinspace\negthinspace= \negthinspace\negthinspace \rho^{(2)}(\vec{r}z,\vec{r}'z';L) \negthinspace/ \negthinspace\rho^{(1)}(\vec{r}z;L)\rho^{(1)}(\vec{r}'z';L)\\
=&\tilde\rho^{(2)}(\vec{r}z,\vec{r}'z';L)/\tilde\rho^{(1)}(\vec{r}z;L)\tilde\rho^{(1)}(\vec{r}'z';L),
\end{align}
where have used Eqs.~\eqref{eq:mpdensity} and~\eqref{eq:mpdensity-a}.

A further measure of structural properties is the density-density correlation function $G(\vec{r}-\vec{r}',z,z';L)$, which
can be decomposed into a self part
\begin{equation}\label{eq:Gs}
G^{(s)}(\vec{r},z,z';L)=\frac{1}{n_{0}}\rho^{(1)}(\vec{r}z;L) \delta(\vec{r})\delta(z-z'),
\end{equation}
and a distinct part
\begin{align}\label{eq:distinct}
&G^{(d)}(\vec{r},z,z';L)\nonumber\\
=& \frac{1}{n_{0}} \Big[\rho^{(2)}(\vec{r}z,\vec{r}'z';L)-
\rho^{(1)}(\vec{r}z;L)\rho^{(1)}(\vec{r}'z';L)\Big],
\end{align}
i.e.
\begin{align}\label{eq:dcorrelator}
G(\vec{r}-\vec{r}',z,z';L)= &G^{(s)}(\vec{r}-\vec{r}',z,z';L)\nonumber\\
 &+ G^{(d)}(\vec{r}-\vec{r}',z,z';L).
\end{align}
The total correlation function, $h(\vec{r},z,z';L)=g(\vec{r},z,z';L)-1$, and the direct correlation function $c(\vec{r},z,z';L)$
are related by the Ornstein-Zernike equation of inhomogeneous fluids~\cite{Hansen:Theory_of_Simple_Liquids,Henderson:Fundamentals_of_inhomogeneous_fluids}
\begin{align}\label{eq:OZreal}
c(\vec{r},z,z')=&h(\vec{r},z,z')  \nonumber \\
&- \int\!\! \diff^2 r''\!\! \int\!\! \diff z'' c(\vec{r}-\vec{r}'',z,z'')n(z'')\nonumber \\
&\times h(\vec{r}''-\vec{r}',z'',z').
\end{align}
It is useful to represent real space functions $f(\vec{r},z)$ as superpositions of symmetry-adapted Fourier modes ~\cite{Lang:2012}
\begin{equation}\label{eq:decom}
f_{\mu}(\vec{q};L)=\int \diff ^2r \diff z f(\vec{r},z;L) \exp(i Q_\mu z) \, \text{e}^{i \vec{q} \cdot \vec{r}},
\end{equation}
where the discrete wave numbers  $Q_{\mu}=2\pi \mu /L$, $\mu \in \mathbb{Z}$ characterize the modulations perpendicular to the walls. Here, $\vec{q}=(q_{x},q_{y})$ are the conventional discrete (for finite $A$) wave vectors in the $x-y$-plane.
Similarly, for real-space correlation functions $F(\vec{r}-\vec{r}',z,z';L)$ we employ the Fourier modes
\begin{align}\label{eq:corrdecom}
F_{\mu \nu}(q;L) =&\int\!\! \diff ^2 r \diff z \diff z'  F(\vec{r},z,z';L)  \nonumber \\
&\times \exp(-i Q_\mu z) \exp(i Q_\nu z') \, \text{e}^{-i \vec{q} \cdot \vec{r}}.
\end{align}
Note, $f_{\mu}(\vec{q};L)$ and $F_{\mu \nu}(q;L)$ depend only on the magnitude $q = | \vec{q}|$ due to the rotational symmetry with respect to the $z$-axis.
The density profile within the slit is decomposed into the discrete wave-numbers
\begin{equation}\label{eq:densitymode}
 n_\mu(L) = \int \diff z \exp( i Q_\mu z) n(z;L).
\end{equation}
The Fourier transform of the density-density correlation function  $G(\vec{r},z,z';L)$ is referred to as generalized structure factor $S_{\mu\nu}(q;L)$ with corresponding decomposition into self and distinct part $S_{\mu \nu}(q;L)=S^{(s)}_{\mu \nu}(q;L)+S^{(d)}_{\mu \nu}(q;L)$. Explicitly one infers

\begin{align}
S_{\mu\nu}^{(s)}(q;L)=& n_{\nu-\mu}(L)/n_{0} \label{eq:self_structure}, \\
S_{\mu\nu}^{(d)}(q;L)=& \int\diff ^2 r  \diff z \diff z' G^{(d)}(\vec{r},z,z';L)  \nonumber \\
&\times \exp(-i Q_\mu z) \exp(i Q_\nu z') \, \text{e}^{-i \vec{q} \cdot \vec{r}}.
\end{align}

The transform  $c_{\mu \nu}(q;L)$ of the direct correlation functions is related to $S_{\mu \nu}(q;L)$ via the inhomogeneous Ornstein-Zernike equation~\cite{Henderson:Fundamentals_of_inhomogeneous_fluids}, which we reformulate in terms of the symmetry-adapted modes~\cite{Lang:2012}
\begin{equation}\label{eq:OZ}
 \mathbf{S}^{-1}(q;L)=\frac{n_{0}}{L^2}\left[\mathbf{v}(L)-\mathbf{c}(q;L)\right],
\end{equation}
with the following matrix notation $[\mathbf{c}(q;L)]_{\mu\nu}=c_{\mu\nu}(q;L)$ etc.  The Fourier modes $v_{\mu \nu}(L)\equiv v_{\nu-\mu}(L)$ of the local volume $v(z)=1/n(z)$ are related to $n_{\mu \nu}(L)\equiv n_{\nu-\mu}(L)$ by
\begin{equation}\label{eq:id}
\mathbf{v}(L)\mathbf{n}(L)=L^2 \mathbf{1},
\end{equation}
see Ref.~\cite{Lang:2010,Lang:2012}.

\section{Two-dimensional limit}\label{Sec:2DLimit}

In the first part of this section we discuss conditions on the interactions such that the $m$-particle density and related correlation functions
 converge for $L\to 0$ properly to their corresponding 2D counterparts. Then, we elaborate a systematic expansion of the structural quantities as introduced in Sec.~\ref{Sec:Basic}.
To avoid cumbersome notation, quantities depending only on the 2D vectors $\vec{r}$ or $\vec{q}$ will be considered as obtained from a canonical average with total interaction potential $V_0(\{\vec{r}\}) \equiv V_0(\{\vec{r} \} ,\{0\})$. For instance $\rho^{(m)}(\vec{r}_1,...,\vec{r}_m)$ denotes the 2D $m$-particle density and
\begin{align}\label{eq:Gd2d}
 &G(\vec{r})= G^{(s)}(\vec{r})+G^{(d)}(\vec{r}), \\
 &G^{(s)}(\vec{r})=\frac{1}{n_{0}}\rho^{(1)}(\vec{r})\delta(\vec{r}) \equiv \delta(\vec{r}), \\
&G^{(d)}(\vec{r}-\vec{r}')=\frac{1}{n_0}\left[\rho^{(2)}(\vec{r},\vec{r}')-\rho^{(1)}(\vec{r})\rho^{(1)}(\vec{r}')\right],
\end{align}
refers to the  density-density correlation function of the corresponding 2D fluid decomposed into the self and distinct part. Of course, $G(\vec{r})$, $G^{(s)}(\vec{r})$, and $G^{(d)}(\vec{r})$ depend on $|\vec{r}|=r$, only.
The planar total correlation function $h(r)$ is connected to the distinct part of the density-density correlation function by $G^{(d)}(r)=n_{0}h(r)$ and the 2D pair-distribution function follows by the relation $g(r)=1+ h(r)$.

\subsection{Existence of the two-dimensional limit}\label{Sec:ex}

The static structure of the confined fluid converges properly to the 2D limit if the fluid becomes structureless in the transverse direction. As will be demonstrated below this requires certain conditions on the microscopic interactions of the particles with the walls, i.e. the approach to the planar limit depends on qualitative features of ${\cal U}_{\pm}(z)$ which determine the wall potential ${\cal U}(z;L)$ (c.f. Eq.~\eqref{eq:potential_wall}). To discuss convergence
with respect to  different wall separations of functions $f(\vec{r}_1z_1,\vec{r}_2z_2,...;L)$ defined on the slit
it will be convenient to establish the dimensionless transversal position  $\tilde{z} = z/L$,  with the fixed domain $\tilde{z} \in [-1/2,1/2]$ as fundamental variable. Then, this function has a proper 2D limit if $f(\vec{r}_1\tilde{z}_1L,\vec{r}_2\tilde{z}_2L,...;L)$ becomes independent of $\tilde{z}_i$ for $L\to 0$. As an example we consider the density profile. From Eq.~\eqref{eq:densitymode} we obtain
\begin{align}\label{eq:rescaledens}
n_{\mu}(L)&=  L\int_{-1/2}^{1/2} \diff \tilde{z} \, n(\tilde{z} L ;L)\exp(2 \pi i \mu \tilde{z}).
\end{align}
The proper 2D limit requires that all non-trivial modes vanish, $n_{\mu}(L)\to n_{0}\delta_{\mu 0}$ for $L\to 0$ which is  satisfied if
 $n(\tilde{z}L;L)\to n_{0}/L$, i.e. the density profile for $L\to 0$  becomes independent of the scaled transversal coordinate $\tilde{z}$. For a hard-sphere liquid with neutral walls this is demonstrated in Fig. 1  of Ref.~\cite{Antonchenko:1984}. This convergence describes a fluid for $L\to 0$, which is determined by the lateral interaction potential $V_0(\{\vec{r}\})$, only.

Since the $m$-particle density $\rho^{(m)}(\vec{r}_1z_1,...,\vec{r}_mz_m;L)$ is entirely determined by the densities $\rho^{(1)}_{\perp}(z;L)$  and $\tilde\rho^{(m)}(\vec{r}_1z_1,...,\vec{r}_mz_m;L)$ (c.f. Eq.~\eqref{eq:mpdensity}), the convergence to the 2D limit requires both densities to  become independent of $\tilde{z}$  and $\tilde{z}_{1},...,\tilde{z}_{m}$, respectively.
As will be demonstrated in subsection ~\ref{Sec:mparticle}, the reduced $m$-particle density $\tilde{\rho}^{(m)}(\vec{r}_1\tilde{z}_1L,...,\vec{r}_m\tilde{z}_mL;L)$ converges properly for $L\to 0$, which is not necessarily the case for $\rho^{(1)}_{\perp}(\tilde{z}L;L)$. Since $\rho^{(1)}_{\perp}(\tilde{z}L;L)$
depends only on ${\cal U}(\tilde{z}L;L)$, its convergence is solely controlled by the particle-wall interaction.

 Assume that the wall potential for fixed $\tilde{z}$ fulfills the smoothness criterion
\begin{equation}
{\cal U}(z= \tilde{z} L;L)-{\cal U}(0;L) = {\cal O}(L),
\label{eq:smoothness_criterion}
\end{equation}
which is valid for potentials that are analytic in $z$.
Then the 'bare' density profile  $\rho^{(1)}_{\perp}(z;L)$ becomes flat even on the scale of the plate distance
\begin{equation}\label{eq:nv}
\rho^{(1)}_{\perp}(\tilde{z}L;L)  =  \frac{1}{L}\left[1 + {\cal O}(L)\right].
\end{equation}
This in turn implies convergence of the Fourier modes of the density profile
\begin{align}
&n_{\mu}(L)= n_{0}[ \delta_{\mu 0}+(1-\delta_{\mu 0}){\cal O}(L) ].
\end{align}
Note, that the normalization of  $n(z;L)$ implies $n_{\mu=0}(L)=N/A=n_{0}$ for all $L$. Thus the property that only the zero Fourier mode of the density profile ($1$-particle density) survives in the limit of $L\to 0$ serves as a \emph{definition} of proper convergence to a two-dimensional fluid.

It is instructive to give a counterexample for a wall potential such that the smoothness criterion is violated and  $n_{\mu}(L)$ does not converge to  $n_{0}\delta_{\mu 0}$.
For instance for symmetric and repulsive walls with  $\mathcal{U}_{\pm}(z)\equiv{\cal U}_{\text{w}}(z) =  a  z^{-\alpha}, a>0, \alpha>0$,  the Boltzmann factor becomes
\begin{align}
&\text{e}^{-\beta [{ \cal U}(\tilde{z} L;L)-{\cal U}(0;L) ]}   = \nonumber \\
&= \exp\{ - \beta a (L/2)^{-\alpha} [ (1-2 \tilde{z})^{-\alpha} + (1+ 2\tilde{z})^{-\alpha} -2 ] \} \nonumber \\
&\to
\begin{cases}
 1 & \text{for } \tilde{z} =0 \\
0 & \text{else}
\end{cases}
\end{align}
as $L\to 0$.
Hence the density profile $n(z=\tilde{z} L;L) \to 0$  for $\tilde{z} \neq 0$, yet by normalization $\int n(z;L) \diff z = n_{0}$. Thus we have demonstrated that in this case the  density profile becomes singular
\begin{equation}
n(z=\tilde{z} L;L) \to (n_0/L) \delta(\tilde{z}),
\end{equation}
in strong contrast to Eq.~\eqref{eq:nv}. Equivalently, the Fourier modes of  the density profile converge as $n_{\mu}(L) \to n_{0}$ for \emph{all} $\mu$. Therefore the density  profile and consequently as well the $m$-particle density  do not have a proper 2D limit for this kind of wall potentials. Note that, e.g. Lennard-Jones and Coulomb potentials belong to the class of wall potentials, for which no proper convergence can be achieved.

\subsection{The $m$-particle density and the density profile}\label{Sec:mparticle}

In this subsection we elaborate an expansion of the $m$-particle density with respect to the wall separation $L$. First, we derive asymptotically exact results for smooth pair potentials, followed by an outline of the corresponding result in the case of hard-core interactions. The $1$-particle density, i.e. the density profile, the most basic quantity characterizing inhomogeneous fluids, will be discussed in more detail.

In Ref. ~\cite{Franosch:2012} it has been demonstrated that the thermodynamic behavior of quasi-two-dimensional fluids can be obtained from a systematic expansion around a 2D reference fluid. This is based on the fact that the transverse coordinates $z_i$ are of order $L$. For \emph{smooth} pair potentials we use the expansion
\begin{equation}\label{eq:distance}
|\vec{x}_i-\vec{x}_j| = r_{ij} + z^2_{ij}/ 2r_{ij}+ {\cal O}(z_{ij})^4,
\end{equation}
with $r_{ij}= |\vec{r}_i - \vec{r}_j|$ and $z_{ij}=(z_i-z_j)$. Then we obtain from Eq.~\eqref{eq:pairpotential}
\begin{align}\label{eq:seriespot}
\exp[-\beta V_0 (\{\vec{x}\})]=&\exp[-\beta V_0 (\{\vec{r}\})] \nonumber\\
&\times\left[1+ \sum \limits _{l=1} ^\infty \sum \limits _{i < j} v_l(\vec{r}_i z_i,\vec{r}_j z_j)\right],
\end{align}
where
\begin{align}\label{eq:corrpotl}
 v_l (\vec{r}_i z_i,\vec{r}_jz_j)=
\begin{cases}
-\beta {\cal V}'(r_{ij})z_{ij}^{2}/2r_{ij} ={\cal O}(L^2)  & \text{if $l =1$},  \\
{\cal O}(L^{2l}) & \text{if $l >1$},
\end{cases}
\end{align}
i.e. $v_l ={\cal O}(L^{2 l}) $ for $l \geq 1$. Note, that the explicit expressions for $v_l(\vec{r}_i z_i,\vec{r}_j z_j)$ become more and more involved with increasing $l$, since Eq.~\eqref{eq:distance} has to be extended up to order $(z_{ij})^{2 l}$.

Here, a comment is in order. If the particles are charged such that they are interacting via Coulomb forces one might be tempted to use for the corresponding interaction potential of the 2D reference fluid the 2D Coulomb potential: $\sim \ln{r_{ij}}$. This does not apply here, since the confined fluid even in the 2D limit is embedded in 3D space. Accordingly, the corresponding 2D potential of the reference fluid in this case is the conventional Coulomb potential proportional to $1/r_{ij}$.

 In the following we restrict the expansion of the $m$-particle density to first order in $L^2$. In this case the calculations simplify due to the factorization of $v_1(\vec{r}_i z_i,\vec{r}_jz_j)$:

\begin{eqnarray}\label{eq:factv1}
v_1(\vec{r}_i z_i,\vec{r}_j z_j)&=&
v_1^\parallel(\vec{r}_i,\vec{r}_j)v_1^\perp(z_i,z_j), \nonumber\\
v_1^\parallel(\vec{r}_i,\vec{r}_j)
&=&- \beta \mathcal{V}'(r_{ij})/2r_{ij},\\
v_1^\perp (z_i,z_j)&=&(z_{ij})^2.\nonumber
\end{eqnarray}

The calculation of higher-order terms is straightforward, but cumbersome.
 Since $\rho_\perp ^{(m)}(z_1,\ldots,z_m;L)$ is explicitly known a priori for a given particle-wall interaction $\mathcal{U}(z;L)$, the expansion has to be carried out for the
reduced $m$-particle density $\tilde{\rho}^{(m)}(\vec{r}_1,z_1,\ldots,\vec{r}_mz_m;L)$, only.
Equations ~\eqref{eq:mpdensity-b} and~\eqref{eq:seriespot} lead to
\begin{equation}\label{eq:seriesmpdensity}
 \tilde{\rho}^{(m)}(\vec{r}_1z_1,\ldots,\vec{r}_mz_m;L) = \sum \limits _{l =0}^\infty \tilde{\rho}^{(m)}_l(\vec{r}_1z_1,\ldots,\vec{r}_mz_m;L),
\end{equation}
where $\tilde{\rho}_l^{(m)}=\mathcal{O}(L^{2 l})$. The explicit evaluation of the leading order and its first correction has been transferred to Appendix~\ref{Sec:acum}. Employing Eq.~\eqref{eq:factv1} and defining averages with respect to the perpendicular ensemble $\langle f(z_{1},...,z_{k}) \rangle_\perp := \int dz_{1}... \int dz_{k} f(z_{1},...,z_{k})\rho_\perp ^{(1)}(z_{1} )...\rho_\perp ^{(1)}(z_{k})$  we obtain from Eq.~\eqref{eq:A13}
to leading order the 2D $m$-particle density
\begin{equation}\label{eq:correctmpdensity-0}
\tilde{\rho}^{(m)}_0(\vec{r}_1z_1,\ldots,\vec{r}_mz_m;L)=
\rho^{(m)}_\parallel(\vec{r}_1,\ldots,\vec{r}_m),
\end{equation}
and  for the leading correction
\begin{align}\label{eq:correctmpdensity-1}
&\tilde{\rho}^{(m)}_1(\vec{r}_1z_1,\ldots,\vec{r}_m z_m;L)\nonumber \\
=& \sum \limits _{1 \leq i <j \leq m} v_{1}^{\perp} (z_i,z_j) v_1^{\parallel} (\vec{r}_i,\vec{r}_j) \rho_{\parallel} ^{(m)}(\vec{r}_1,\ldots, \vec{r}_m)\nonumber \\
& + \negthinspace\negthinspace \sum \limits _{i=1}^m \! \langle v_1^{\perp} (z_i,z_{m+1})\rangle_\perp \! \! \int \! \! d^2r_{m+1}v_1^\parallel (\vec{r}_{i}, \vec{r}_{m+1}) \rho_{\parallel} ^{(m+1)}(\vec{r}_1,\ldots, \vec{r}_{m+1}) \nonumber \\
&+ \frac 1 2 \langle v_1^\perp (z_{m+1}, z_{m+2})\rangle _\perp  \int d^2r_{m+1} \int d^2r_{m+2} v_1^\parallel (\vec{r}_{m+1},\vec{r}_{m+2}) \nonumber \\
&\times \left[\rho _\parallel ^{(m+2)} (\vec{r}_1,\ldots, \vec{r}_{m+2})-\rho_\parallel ^{(m)}(\vec{r}_1,\ldots, \vec{r}_m)\rho _\parallel ^{(2)}(\vec{r}_{m+1},\vec{r}_{m+2})\right]\nonumber \\
&- \langle v_1^\perp (z_{m+1},z_{m+2})\rangle _\perp  \left(\frac{n_0^2 \kappa_T}{2  \beta}\right) \left(\frac {\partial}{\partial n_0} \Big|_T \rho^{(m)}_\parallel (\vec{r}_1,\ldots, \vec{r}_m)\right) \nonumber\\
&\times \int d^2r_{m+2} v_1^\parallel (\vec{r}_{m+1}, \vec{r}_{m+2}) \frac{\partial}{\partial n_0}\Big|_T \rho_\parallel ^{(2)} (\vec{r}_{m+1},\vec{r}_{m+2}).
\end{align}
The last term in Eq.~\eqref{eq:correctmpdensity-1} results from the thermodynamic limit (cf. Ref.~\cite{Smith:1971}). Its integral term does not depend on $\vec{r}_{m+1}$ due to lateral translational invariance. Here, $\kappa_T=\left[n_{0}\partial \Sigma / \partial n_{0}\Big|_{T} \right]^{-1}$ refers to the 2D isothermal compressibility of the fluid with surface tension given by $\Sigma=n_{0}k_{B}T+ k_{B}T\partial \ln (Z_{\parallel})/ \partial A \Big|_{T,N}$. The density derivatives of $\rho_\parallel ^{(2)}$ and $\rho_\parallel^{(m)}$ can be expressed by $\rho_\parallel ^{(k)}$~\cite{Schofield:1966}

\begin{align}\label{eq:densityderiv}
\frac {\partial} {\partial n_0}\Big|_T &\rho_\parallel ^{(k)} (\vec{r}_1,\ldots,\vec{r}_k)\nonumber \\
=& \Big\{k \rho_\parallel ^{(k)}(\vec{r}_1,\ldots,\vec{r}_k) +\int d^2r_{k+1} \big[ \rho_\parallel ^{(k+1)} (\vec{r}_1,\ldots,\vec{r}_{k+1})\nonumber\\
&- \rho_\parallel ^{(k)}(\vec{r}_1,\ldots,\vec{r}_k) \rho_\parallel ^{(1)}(\vec{r}_{k+1})\big]\Big\}/(n_0^2\beta ^{-1} \kappa_T).
\end{align}
Additionally, one can prove
\begin{align}\label{eq:factmpdensity}
&\tilde{\rho}_1^{(m)} (\vec{r} _1z_1,\ldots,\vec{r}_mz_m;L)\nonumber\\
&\rightarrow \tilde{\rho}_0^{(k)} (\vec{r}_1z_1,\ldots, \vec{r}_kz_k;L) \tilde{\rho}_1^{(m-k)} (\vec{r}_{k+1}z_{k+1},\ldots,\vec{r}_mz_m;L) \nonumber \\
&+ \tilde{\rho}_1^{(k)}(\vec{r}_1z_1,\ldots,\vec{r}_kz_k;L) \tilde{\rho}_0^{(m-k)} (\vec{r}_{k+1}z_{k+1},\ldots,\vec{r}_mz_m;L),
\end{align}
for $\vec{r}_i,i=1,\ldots, k$ fixed and $|\vec{r}_j| \rightarrow \infty, \; j=k+1,\ldots,m$. This property preserves the factorization of the $m$-particle density~\cite{Lebowitz:1961} to first order if a subset of particles $(1,...,k)$ is fixed and the remaining subset $(k+1,...,m)$ is moved to infinity. Further, the result of Eq.~\eqref{eq:correctmpdensity-1} formally has the same structure as the first-order correction of the pair-distribution function of a bulk fluid perturbed by an additional pair potential~\cite{Smith:1971}. However, Eq.~\eqref{eq:correctmpdensity-1} is valid for a 2D fluid perturbed by the translational d.o.f. and holds for all $m$.

In the case of hard-sphere interactions a corresponding expansion can be performed by utilizing the cluster expansion derived in~\cite{Franosch:2012}. There, a 2D liquid of hard disks of reduced diameter $\sigma _L =(\sigma ^2-L^2)^{1/2}$ has been chosen as reference fluid and the cluster function

\begin{align}\label{eq:clusterf}
f_{ij}&\equiv f(r_{ij}, z_i,z_j)\nonumber\\ 
&= \Theta [r^2_{ij} + (z_i-z_j)^2-\sigma ^2] - \Theta (r^2_{ij}-\sigma ^2_L),
\end{align}
has been introduced, where $\Theta (x)$ refers to the Heaviside function. The cluster function satisfies the identity~\cite{Franosch:2012}

\begin{align}\label{eq:clusterexp}
\exp[-\beta V_0 (\{\vec{x}\})]& = \exp [-\beta W_0(\{\vec{r}\})]  \prod \limits _{i<j}(1+ f_{ij}) \nonumber \\
&=\exp[-\beta W_0(\{\vec{r}\})] \left[1+\sum \limits _{i<j} f_{ij}+ \ldots\right],
\end{align}
where $\exp [-\beta W_0(\{\vec{r}\})]= \prod \limits _{i<j} \Theta (r_{ij}^2 - \sigma ^2_L)$ is the Boltzmann factor of the hard disk fluid. Since
the first-order term $\sum \limits _{i<j} f_{ij}$ corresponds to $\sum \limits _{i<j} v_1(\vec{r}_i z_i,\vec{r}_jz_j)$
in Eq.~\eqref{eq:seriespot}  one obtains $\tilde{\rho}^{(m)}$ for hard-core interactions
from Eq.~\eqref{eq:A7} by the replacement $v_1\rightarrow f$. The integrations over
$z_i,z_j$ and $\vec{r}_i$, $\vec{r}_j$ can still be disentangled, see Appendix~\ref{Sec:ccluster} for an example.
As a result the $m$-particle density $\tilde {\rho}^{(m)}$ for $m=0,1$ is expressed by $\rho_\parallel ^{(m)}$, the $m$-particle density of hard disks with reduced diameter $\sigma _L$. To eliminate the dependence of $L$ of the reference fluid,  a further expansion with respect to $L$ is required in order to obtain the corrections with respect to a 2D-reference fluid with diameter $\sigma$.
Beyond the first order one can not follow the expansion, Eq.~\eqref{eq:seriespot}, by a simple replacement,
but it becomes necessary to perform a cluster expansion as elaborated in Ref.~\cite{Franosch:2012}.
Restricting the expansion to zero and first-order corrections allows to determine
$\tilde{\rho}^{(m)}, \; m = 0,1$ directly from Eq.~\eqref{eq:correctmpdensity-1}
by choosing for $\mathcal{V}(r_{ij})$ the hard disk potential with a  hard-core diameter $\sigma$,
replacing $-[\beta\mathcal{V}'(r_{ij})/2r_{ij}] \rho_\parallel ^{(m)} (\ldots, \vec{r}_i,\ldots,\vec{r}_j,\ldots)$
by $(\exp[\beta \mathcal{V}(r_{ij})]/2 r_{ij})\rho_\parallel^{(m)}(\ldots, \vec{r}_i,\ldots,\vec{r}_j,\ldots)\diff (\exp [-\beta \mathcal{V}(r_{ij})])/\diff r_{ij}$
and subsequently taking the limit $\lim_{r_{ij}\rightarrow \sigma ^+}\diff \exp [-\beta \mathcal{V}(r_{ij})]/\diff r_{ij}=\delta (r_{ij}-\sigma^{+})$. This strategy is demonstrated below for the density profile.
The first-order correction, Eq.~\eqref{eq:correctmpdensity-1}, strongly simplifies for $m=1$ (see Appendix ~\ref{Sec:bdensity}). One obtains with $\rho^{(2)}_{\parallel}(r)=n_{0}^2g(r)$ from Eq.~\eqref{eq:B5}
\begin{align}\label{eq:profile-1}
n(z;L) \equiv& \rho^{(1)} (\vec{r}z;L)\nonumber\\
=& n_0 \rho_\perp ^{(1)} (z;L)\Big\{1+\left[\langle v_1^\perp(z,z_2)\rangle _\perp - \langle v_1^\perp (z_2,z_3)\rangle _\perp \right]\nonumber \\
&\thinspace \times n_0 \int d^2r' v_1^\parallel (0,\vec{r}')g(r') + \mathcal{O} (L^4)\Big\},
\end{align}
with $g(r)$ the pair-distribution function of the 2D reference fluid.
Note that the average $\langle v_1^\perp(z,z_2)\rangle _\perp$ is taken  with respect to $z_2$, only. Since $\int dz \rho_\perp ^{(1)} (z;L) [\langle v_1^\perp (z,z_2)\rangle _\perp - \langle v_1^\perp (z_2,z_3)\rangle ] =0$ the first-order correction does not contribute to the normalization $\int \diff z n(z;L) = n_{0}$. Substituting $v_{1}^{\perp}$  and $v_{1}^{\parallel}$ from Eq.~\eqref{eq:factv1} into Eq.~\eqref{eq:profile-1} one obtains the explicit $z$-dependence
\begin{align}\label{eq:climit}
n(z;L) = &  n_0  \rho^{(1)}_{\perp}(z;L)\Big[1+ \pi n_{0} C \Big\{ z^2-2z\langle z_{1} \rangle_{\perp}+ \langle z_{1}^2 \rangle_{\perp}\nonumber\\
&-\langle(z_{1}-z_{2})^2\rangle_{\perp} \Big\}+\mathcal{O}(L^4)\Big],
\end{align}
where the prefactor
\begin{equation}\label{eq:prefactorprof-1}
C= -\beta \int_0^\infty \diff r \mathcal{V}'(r)g(r),
\end{equation}
characterizes the first-order corrections. Since the curly bracket in Eq.~\eqref{eq:climit} is of order $\mathcal{O}(L^2)$ the correction is proportional to $n_0 L^2$, the smallness parameter identified in Ref.~\cite{Franosch:2012}.

Following the procedure as explained above the prefactor $C$ becomes in the case of pure hard-core interactions
\begin{equation}\label{eq:prefactorprof-2}
C=g(\sigma^{+}).
\end{equation}
If the potentials  ${\cal U}_{\pm}(z)$ are analytic at $z=0$ we obtain from Eq.~\eqref{eq:mpdensity-a}
\begin{align}\label{eq:seriesrhoperp}
\rho^{(1)}_{\perp}(z;L)&= \frac{1}{L}\Big\{1-\beta\left({\cal U}_{-}'(0)-{\cal U}_{+}'(0)\right)z+ \mathcal{O}(Lz) +\mathcal{O}(z^2)\Big\}.
\end{align}
In the case of \emph{symmetric} walls, i.e. ${\cal U}_{+}(z) \equiv {\cal U}_{-}(z) $, it is $\langle z_{1} \rangle_{\perp} =0$ and it follows from Eqs.~\eqref{eq:climit} and  ~\eqref{eq:seriesrhoperp} that the profile is parabolic.  Consequently, the first-order correction of the profile is of order $\mathcal{O}(L^2)$.
For \emph{asymmetric} walls and $L\to 0$ the density $n(z;L)$  is dominated by the bare profile $\rho^{(1)}_{\perp}(z;L)$, which is linear, i.e. the leading order correction is of order $\mathcal{O}(L)$. Hence, the first-order correction depends qualitatively on the symmetry of the wall-potential. This is as well reflected in the Fourier modes of the density profile
\begin{align}\label{eq:corrpotln}
n_{\mu}(L)=
\begin{cases}
n_{0}=N/A   & \text{for $\mu=0$},  \\
{\cal O}(L^{\gamma}) & \text{else},
\end{cases}
\end{align}

and of the local volume
\begin{align}\label{eq:vmu}
v_{\mu}(L)=\frac{L^2}{n_{0}} 
\begin{cases}
[1+\mathcal{O}(L^{2\gamma})]& \text{for $\mu=0$},  \\
\mathcal{O}(L^{\gamma}) & \text{else},
\end{cases}
\end{align}
where $\gamma=1$ for asymmetric and $\gamma=2$ for symmetric walls.

A special case  of symmetric walls are neutral walls, i.e. ${\cal U}_{+}(z)={\cal U}_{-}(z) \equiv 0$.
In this case  Eq.~\eqref{eq:climit} reduces to
\begin{equation}\label{eq:profile_parabolic}
n(z;L)= \frac{n_0}{L} \left\{1+ \pi (n_0 L^2)  C
\left[ \left(\frac{z}{L}\right)^2-\frac{1}{12}\right] +  {\cal O}(n_0 L^2)^2 \right\}.
\end{equation}
In the case of hard spheres, $C=g(\sigma^{+})$, up to order $\mathcal{O}(L^2)$ the density profile, Eq.~\eqref{eq:profile_parabolic}, satisfies the contact theorem $n(z=\pm L/2)=p/k_{B}T$, see e.g.~\cite{Henderson:1979,Roth:2010}, with the transversal pressure $p=k_{B}T n_{0}/L+ k_{B}T A^{-1} \partial \ln{Z} / {\partial L} \Big|_{N,A}$ exerted on the walls, which has been evaluated recently ~\cite{Franosch:2012} to
\begin{align}
p&=n_{0}k_{B}T/L\left[1+\frac{1}{6} \pi (n_{0} L^2) g(\sigma^+)+\mathcal{O}(L^4)\right]\nonumber\\
&\equiv k_{B}T n(z=\pm L/2;L).
\end{align}

\subsection{Correlation functions and their Fourier transforms}

The various two-point correlation functions introduced in Sec.~\ref{Sec:Basic} involve the $1$-particle density
$\rho^{(1)}(\vec{r}z;L)$ from Eq.~\eqref{eq:profile-1} and the $2$-particle density
$\rho^{(2)}(\vec{r}z,\vec{r}'z';L)$. From Eqs.~\eqref{eq:mpdensity} and ~\eqref{eq:correctmpdensity-1} one obtains for $m=2$ the expansion

\begin{align}\label{eq:correct2pdensity}
\rho&^{(2)}(\vec{r}z,\vec{r}'z';L)\nonumber \\
=&\rho^{(1)}_{\perp}(z;L)\rho^{(1)}_{\perp}(z';L) \nonumber\\
&\times \Big\{\left[1+v_{1}^{\perp}(z,z')v_{1}^{\parallel}(\vec{r},\vec{r}') \right]\rho_{\parallel}^{(2)}(\vec{r},\vec{r}') \nonumber\\
&+\negthinspace \negthinspace\left[\langle v_1^\perp(z,z_3)\rangle _\perp + \langle v_1^\perp (z',z_3)\rangle _\perp \right]\negthinspace \negthinspace \negthinspace\int\negthinspace  \negthinspace d^2r_{3}v_1^\parallel (\vec{r}, \vec{r}_{3}) \rho_{\parallel} ^{(3)}(\vec{r},\vec{r}',\vec{r}_{3}) \nonumber\\
& + \frac 1 2 \langle v_1^\perp (z_{3}, z_{4})\rangle _\perp  \int d^2r_{3} \int d^2r_{4} v_1^\parallel (\vec{r}_{3},\vec{r}_{4})\nonumber \\
&\times\left[\rho _\parallel ^{(4)} (\vec{r},\vec{r}',\vec{r}_{3},\vec{r}_{4})- \rho_\parallel ^{(2)}(\vec{r},\vec{r}')\rho_\parallel ^{(2)}(\vec{r}_{3},\vec{r}_{4})\right] \nonumber\\
&- \langle v_1^\perp (z_{3},z_{4})\rangle _\perp \left( \frac{n_0^2 \kappa_T}{2 \beta}\right) \left(\frac {\partial}{\partial n_0} \Big|_T \rho^{(2)}_\parallel (\vec{r}, \vec{r}')\right)\nonumber\\
&\times\int d^2r_{4} v_1^\parallel (\vec{r}_{3}, \vec{r}_{4}) \frac{\partial}{\partial n_0}\Big|_T \rho_\parallel ^{(2)} (\vec{r}_{3},\vec{r}_{4}) + \mathcal{O}(L^4)\Big\}.
\end{align}

The leading order factorizes to $\rho^{(1)}_{\perp}\negthinspace(z;L)\rho^{(1)}_{\perp}\negthinspace (z';L) \rho_{\parallel}^{(2)}\negthinspace(\vec{r},\vec{r}')\negthinspace \negthinspace=\negthinspace \negthinspace n_{0}^2 \rho^{(1)}_{\perp}(z;L)\rho^{(1)}_{\perp}(z';L)g(|\vec{r}-\vec{r}'|)$, which is a consequence of the decoupling of the lateral and transversal degrees of freedom in the limit $L\to 0$. Note, that this expression is different from the superposition approximation originally suggested in Ref.~\cite{Green:Structure_of_liquids}, where $\rho^{(2)}(\vec{r}z,\vec{r}'z';L)$
is replaced by $\rho^{(1)}(z;L)\rho^{(1)}(z';L) g^{(3D)}(\sqrt{(\vec{r}-\vec{r}')^2+(z-z')^2})=\rho^{(1)}(z;L)\rho^{(1)}(z';L)[g^{(3D)}(|\vec{r}-\vec{r}|)+\mathcal{O}(L^2)]$ with $g^{(3D)}(|\vec{x}|)$ being the pair-distribution  function of the \emph{bulk} fluid.

The first-order correction of $\rho^{(2)}(\vec{r}z,\vec{r}'z';L)$ requires $m$-particle densities $\rho^{(m)}_{\parallel}(\vec{r}_1,...,\vec{r}_m)$ with $m=2,3$ and $4$ of the 2D reference fluid. These correlation functions can be determined either by computer simulations or by suitable approximations.
For instance, one can combine integral-equation theory to evaluate $\rho^{(2)}_{\parallel}(\vec{r}_1,\vec{r}_2)$ and then to obtain in superposition approximation $\rho^{(m)}_{\parallel}$ for $m=3$ and $m=4$. With Eqs.~\eqref{eq:profile-1} and ~\eqref{eq:correct2pdensity} the first-order correction
$g_{1}(\vec{r}z,\vec{r}'z';L)$ of the pair-distribution function $g(\vec{r}z,\vec{r}'z';L)$ follows from Eq.~\eqref{eq:pairdf}.
Using the factorization property of  Eq.~\eqref{eq:factmpdensity} one can readily prove that the correction vanishes $g_{1}(\vec{r}z,\vec{r}'z';L) \to 0$ for $| \vec{r}-\vec{r}'| \to \infty$.
The $(z,z')$-dependence of $g(\vec{r}z,\vec{r}'z';L)$  is given by $(z-z')^2$,
$\langle (z-z_{3})^2\rangle _\perp$ and  $\langle (z'-z_{3})^2\rangle _\perp$.

Finally we determine the rapidity of convergence of various correlation functions and their Fourier transforms to their respective 2D counterparts.

From Eqs.~\eqref{eq:mpdensity},~\eqref{eq:mpdensity-a},~\eqref{eq:pairdf} and ~\eqref{eq:seriesmpdensity} it follows with $\tilde\rho^{(m)}_{1}=\mathcal{O}(L^2)$ for all $m$
\begin{equation}\label{eq:correctpdf}
g(\vec{r}z,\vec{r}'z';L)= g(r)+\mathcal{O}(L^2),
\end{equation}
independent of the type of the wall potential. The same holds for the total correlation function
\begin{equation}\label{eq:correcttcf}
h(\vec{r}z,\vec{r}'z';L)= h(r)+\mathcal{O}(L^2).
\end{equation}
To discuss the direct correlation function $c(\vec{r}, z,z';L)$ in confined geometry, we employ the inhomogeneous  Ornstein-Zernike equation ~\eqref{eq:OZreal}.
We will prove that $\delta h(\vec{r},z,z';L) = h(\vec{r},z,z';L)-h(r)  = {\cal O}(L^2)$ implies  $\delta c(r,z,z';L) = c(\vec{r},z,z';L)-c(r)  = {\cal O}(L^2)$. Inserting the leading order into Eq.~\eqref{eq:OZreal} and using the sum rule $\int  n(z;L) \diff z=n_0$ one recovers the  homogeneous 2D Ornstein-Zernike equation
\begin{align}\label{eq:OZ2D}
c(\vec{r})=h(\vec{r})- n_{0}\!\int \diff^2 r''c(\vec{r}-\vec{r}'')h(\vec{r}''-\vec{r}').
\end{align}
The next order then constitutes an integral equation for the first correction of the direct correlation function
\begin{align}
&\delta c(r,z,z';L)= \delta h(r,z,z';L)\nonumber\\
&-n_{0} \int\!\! \diff^2 r'' \!\!\int\!\!\diff \tilde{z}'' \delta c(\vec{r}-\vec{r}'',z,z''= \tilde{z}'' L;L)h(\vec{r}''-\vec{r}')\nonumber \\
&-n_{0} \int\!\! \diff^2 r'' \!\!\int\!\! \diff \tilde{z}'' c(\vec{r}-\vec{r}'')\delta h(\vec{r}''-\vec{r}',z''=\tilde{z}'' L,z';L).
\end{align}
In particular to this order the density profile does not contribute and with
$\delta h = \mathcal{O}(L^2)$ we conclude that $\delta c = \mathcal{O}(L^2)$, irrespective of the particle-wall interaction.

The self and distinct part of the  density-density correlation function, $G^{(s)}$ and $G^{(d)}$ from  Eq.~\eqref{eq:Gs} and  Eq.~\eqref{eq:distinct}, respectively, involve $\rho^{(1)}_{\perp}$. For analytic wall potentials we obtain with Eq.~\eqref{eq:seriesrhoperp}
\begin{equation}\label{eq:correctdensitycorrel-s}
G^{(s)}(\vec{r},z,z';L)=\frac{1}{L}[1+ \mathcal{O}(L^{\gamma})]\delta(\vec{r})\delta(z-z'),
\end{equation}
and
\begin{equation}\label{eq:correctdensitycorrel-d}
G^{(d)}(\vec{r},z,z';L)
=\frac{1}{L^2}\left[G^{(d)}(\vec{r})+ \mathcal{O}(L^{\gamma})\right],
\end{equation}
i.e. the first-order correction of $G^{(s)}$ and $G^{(d)}$ depends on the wall type.

The Fourier transforms of all these correlation functions readily follow
\begin{equation}\label{eq: Fourier-g}
 g_{\mu \nu}(q)= L^2 g(q)\left[\delta_{\mu 0} \delta_{\nu 0}+\mathcal{O} (L^2)\right],
\end{equation}
\begin{equation}\label{eq: Fourier-h}
 h_{\mu \nu}(q)= L^2 h(q)\left[\delta_{\mu 0} \delta_{\nu 0}+\mathcal{O} (L^2)\right],
\end{equation}
\begin{equation}\label{eq: Fourier-c}
 c_{\mu \nu}(q)= L^2 c(q)\left[\delta_{\mu 0} \delta_{\nu 0}+\mathcal{O} (L^2)\right].
\end{equation}
We find the leading order for $L\to 0$ is merely given by the $0-0$ component, while the correction and \emph{all} remaining matrix elements vanish as  $\mathcal{O}(L^2)$ irrespective of the wall potential. This stands in contrast to the corresponding Fourier transform of $G^{(s)}$ and $G^{(d)}$
\begin{align}\label{eq:Fourier-Ss}
S_{\mu \nu}^{(s)}(q;L)=
\begin{cases}
1   & \text{for $\mu=\nu$},  \\
 {\cal O}(L^{\gamma}) & \text{else},
\end{cases}
\end{align}
\begin{align}\label{eq:Fourier-Sd}
S_{\mu \nu}^{(d)}(q;L)=
\begin{cases}
S^{(d)}(q)[1+\mathcal{O}(L^2)]   & \text{for $\mu=\nu=0$},  \\
 {\cal O}(L^{\gamma}) & \text{else},
\end{cases}
\end{align}
which yields for the generalized structure factor
\begin{align}\label{eq:Fourier-S}
S_{\mu \nu}(q;L)=
\begin{cases}
S(q)[1+\mathcal{O}(L^2)]   & \text{for $\mu=\nu=0$},  \\
(1-\delta_{\mu 0})\delta_{\mu \nu}+{\cal O}(L^{\gamma}) & \text{else},
\end{cases}
\end{align}
where $S(q)$ refers to the static structure factor of the 2D reference fluid. $S_{\mu \nu}^{(s)}(q;L)$, $S_{\mu \nu}^{(d)}(q;L)$ and $S_{\mu \nu}(q;L)$ have a proper 2D limit only if the criterion (cf.Eq.~\eqref{eq:smoothness_criterion}) is fulfilled. In case of proper convergence these structure factors become diagonal in $\mu$ and $\nu$. In contrast,  since $S_{00}^{(d)}(q;L)=\int\diff ^2 r  \diff z \diff z' n(z)[h(r)+\mathcal{O}(L^2)]n(z') \text{e}^{-i \vec{q} \cdot \vec{r}}/n_{0}=n_0 h(q)[1+\mathcal{O}(L^2)]$ due to the normalization $\int \diff z n(z;L) = n_{0}$, the convergence of the in-planar structure factor $S_{00}(q)$ is of order $\mathcal{O}(L^2)$ and irrespective of the particle-wall interaction.

Finally we mention that  the analytic dependence of $\mathcal{U}(z;L)$ on $z$ is too strict. It is sufficient to require $\mathcal{U}(z=\tilde{z}L;L)-\mathcal{U}(0;L)=o(L^0)$, where the little-$o$ Landau symbol $h(x)=o(x)$  indicates that for $x\to 0$ the function $h(x)$ converges faster to zero than $x$. In this case we have to replace $\mathcal{O}(L)$ by $o(L^0)$ in the estimates and find that the static quantities converge to their respective 2D limit, yet without specification on the rapidity of convergence.

\section{Summary and Conclusions}

For a fluid confined between two parallel walls we have investigated the regime of quasi-two-dimensionality, where the accessible distance $L$ between the adjacent walls becomes small. The focus has been on the behavior of structural quantities, which are entirely described by the $m$-particle density. This quantity can be factorized into a density of transversal d.o.f. depending only on the wall-potential and into a reduced $m$-particle density containing the mutual interactions of the particles in the slit. In the limit $L \to 0$ we analytically determine the $m$-particle density by taking the transversal degrees of freedom as a small perturbation.

To leading order the reduced $m$-particle density is identical to the corresponding $m$-particle density of the 2D reference fluid, which is a consequence of the decoupling mechanism of the transversal and lateral d.o.f. in the 2D limit. The next-to-leading order contains information on the transversal degrees of freedom  $(z_{1},...,z_{m})$, which we find to be quadratic in $z_{i}$ and $z_{j}$. Therefore the leading corrections are of order $L^2$ irrespective of the particular particle-wall interactions. Our analysis has been made explicit for smooth interaction potentials, but we also outline the strategy in the case of hard-core interactions, where a systematic cluster expansion is inevitable~\cite{Franosch:2012}.

The most basic structural entity characterizing the structure of the liquid is the density profile, which merely depends on the transversal position $z$ due to translational symmetry along the walls. Its first-order correction is proportional to $-\beta \int \diff r \mathcal{V}'(r)g(r)$(cf. Eq.~\eqref{eq:profile-1}) with $g(r)$ being the pair-distribution function of the 2D reference fluid. This factor represents a measure of the profile's curvature and therefore of the deviation from flatness. In case of a hard-sphere fluid this proportionality factor becomes $g(\sigma^+)$, the 2D pair-distribution function at contact. This finding differs from  the result in Ref.~\cite{Antonchenko:1984} for the curvature of the density profile. There, the first equation of the Born-Green-Yvon hierarchy for the $2$-particle density has been truncated via the superposition approximation~\cite{Green:Structure_of_liquids}. This superposition approximation involves the pair-distribution function of the 3D \emph{bulk} fluid instead of $g(\sigma^+)$ causing a substantial quantitative mismatch for the curvature. Our result proves, that for $L \to 0$ the superposition principle becomes exact only if the 3D pair-distribution function is replaced by its 2D counterpart.

For the various structural quantities we uncover a hierarchy in terms of the rapidity of convergence. For example, the pair-distribution function $g(\vec{r},z,z')$, which is closely related to the $2$-particle density (cf. Eq. ~\eqref{eq:pairdf})  converges as $L^2$ and its 2D limit always exists. Similarly, this holds for the total and direct correlation function and their rapidity of convergence is independent of the specific type of wall potential.
In contrast, for different structural entities we have found that the 2D limit is rather subtle and depends sensitively on the properties of the wall potentials ${\cal U}_{\pm}(z)$. For instance, for the density profile we have demonstrated that for wall potentials diverging for $z\to 0$, such as for the Lennard-Jones or Coulomb potential, the profile becomes singular for $L\to 0$. The same holds for the density-density correlation function $G(\vec{r}z,\vec{r}'z';L)$. For such wall potentials these quantities do not become flat in the 2D limit. To obtain a proper 2D limit the wall potential $\mathcal{U}(z;L)$ has to fulfill a smoothness criterion as discussed in subsection \ref{Sec:ex}. It is fulfilled if ${\cal U}_{\pm}(z)$ is analytic at $z=0$. However, the condition of analyticity can be weakened. Convergence  also holds provided the wall potential approaches its average value everywhere in the slit.

In both cases, only the zero mode in the Fourier decomposition of the density profile  survives and coincides with the planar density $n_0$. Similarly,
the non-diagonal elements of  the structure factors $S_{\mu\nu}(q;L)$  vanish as the walls approach each other, whereas $S_{00}(q;L)$ converges to $S(q)$, the structure factor of the 2D fluid, and the remaining diagonal elements become unity. While generally a proper convergence of the structure factor depends on the particle-wall interactions, we find that $S_{00}(q;L)$ always converges as $\mathcal{O}(L^2)$ to its 2D counterpart irrespective of the wall potential, which is a manifestation of its sole dependence on the lateral coordinates of the fluid~\cite{Franosch:2012}. In the case of the direct correlation function, we find that the leading order is determined by the matrix element $c_{00}(q;L)$, which is related to the 2D direct correlation function $c(q)$ for $L\to 0$ (cf. Eq.~\eqref{eq: Fourier-c}). The leading correction and  \emph{all} remaining Fourier components $c_{\mu\nu}(q;L)$ converge as $\mathcal{O}(L^2)$ independent of the particle-wall interaction.

Our results provide estimates on the rapidity of convergence. However, the range of validity of the leading order remains unknown in general. In the case of the equilibrium phase diagram for hard spheres of diameter $\sigma$  and neutral walls it has been shown that for $L\lesssim 0.5\sigma$ the leading order describes the phase transition lines quantitatively (cf. figure 3 in Ref.~\cite{Franosch:2012}), i.e. there is almost no influence of the transversal degrees of freedom up to $L\lesssim 0.5\sigma$. We expect  the density profile and the density-density correlation function
to be described quantitatively for similar plate distances.
 Let us compare our predictions with the Monte Carlo results of Ref.~\cite{Antonchenko:1984} for the same system with fixed chemical potential corresponding to a bulk density $n \sigma^3 = 0.5$.
Their figure 1 shows that the profile is practically flat for the smallest wall separation $L = 0.1\sigma$. Consequently, the confined fluid behaves approximately as a 2D fluid. However,
for the next largest value $L = 0.5\sigma$ presented in the same figure  the profile is already parabolic and the subsequent figure  of that paper reveals that the parabolic shape
extends up to $L = 0.8\sigma$. Hence the leading-order correction in $n_0 L^2$ in Eq.~\eqref{eq:profile_parabolic} is sufficient to describe the density profile up to $L = 0.8\sigma$, at least for
the chemical potential chosen to match the bulk density $n \sigma^3 = 0.5$.

The analytical results including leading-order corrections elaborated in this work provide testable predictions, which can be rationalized by computer simulations or experiments. As our results become exact in the limit of small wall separations, they serve as a reference for approximate theories. For instance, in case of hard spheres enclosed between neutral walls we have proven that the curvature of the density profile is determined by the 2D contact value  $g(\sigma^+)$  rather than by $g^{(3D)}(\sigma^+)$ as it has been suggested by the superposition approximation~\cite{Antonchenko:1984}.  Vice versa, our finding enables to measure the contact value of the 2D pair-distribution function via the curvature of the parabolic profile.

\begin{acknowledgments}
We kindly thank Bob Evans for stimulating discussions.
This work has been supported by the
Deutsche Forschungsgemeinschaft DFG via the  Research Unit FOR1394 'Nonlinear response to
probe vitrification'.
S.L.  acknowledges support by the Cluster
of Excellence 'Engineering of Advanced Materials' at the
University of Erlangen-Nuremberg funded by the DFG.
\end{acknowledgments}

\appendix

\section{Calculation of $\tilde{\rho}_l^{(m)} (\vec{r}_1 z_1,\ldots,\vec{r}_m z_m,L)$ for $l = 0,1$}\label{Sec:acum}

We introduce in this Appendix a shorthand notation, where the positions $\vec{x}_j = (\vec{r}_j,z_j)$  are abbreviated by $(j^{\parallel},j^{\perp})= j$ and we adopt the convention
\begin{align}
\int dj (\cdots) &\equiv \int d^\parallel j \int d ^\perp j (\cdots)\nonumber \\
&\equiv \int \limits _A d^2r_j \int \limits _{-L/2}^{L/2} d z_j  \exp [-\beta \mathcal{U} (z_{j};L)](\cdots) /z_{\perp}(L).
\end{align}
Here, $z_\perp(L) = \int \limits _{-L/2}^{L/2} d z \exp [-\beta \mathcal{U} (z;L)]$ refers to the configurational part of the partition function for a single particle interacting with the walls, only.
Functions $f^\parallel (\vec{r}_1,\vec{r}_2,\ldots)$ and $f^\perp(z_1,z_2,\ldots)$ depending on the lateral and transversal d.o.f., respectively, are denoted by  $f^\parallel(1,2,\ldots)$ and $f^\perp(1,2,\ldots)$, so that the superscripts uniquely define the variables.

Substituting Eq.~\eqref{eq:seriespot} into the integral term of Eq.~\eqref{eq:mpdensity-b} and using the decomposition of the interaction potential

\begin{align}\label{eq:A1}
\sum \limits _{1\leq i < j \leq N} v_1 (i,j) =&  \sum \limits _{1\leq i < j \leq m}v_1 (i,j) + \sum \limits _{i=1}^m \sum \limits _{j=m+1}^N v_1(i,j)\nonumber\\
&+ \sum \limits _{m+1 \leq i < j \leq N} v_1(i,j),
\end{align}
one obtains
\begin{align}\label{eq:A2}
&\frac{N!}{(N-m)!} \frac {1}{Z_\parallel} \int \prod \limits _{j=m+1}^N dj \; \exp [-\beta V_0 (1,\ldots,N)]\nonumber\\
=& \frac {N!}{(N-m)!}\frac {1}{Z_\parallel} \int \prod \limits _{j=m+1}^N dj \; \exp [-\beta V_0^\parallel  (1,\ldots,N)] \nonumber \\
&\times \Big{\{} 1 + \sum \limits _{1\leq i< j \leq m} v_1(i,j) + (N-m) \sum \limits_{i=1}^m v_1(i,m+1)\nonumber \\
&+ \frac 1 2 (N-m)(N-m-1) v_1 (m+1, m+2) + \mathcal{O}(L^4) \Big{\}},
\end{align}
where the invariance under relabeling of the summation indices has been employed. The r.h.s. of Eq.~\eqref{eq:A2} can be expressed by the $m$-particle density of the 2D reference fluid
\begin{align}\label{eq:A3}
\rho_\parallel &^{(m)} (1,\ldots,m;N)\nonumber \\
:=& \frac {N!}{(N-m)!} \int \prod \limits _{j=m+1}^N d^\parallel j\; \exp [-\beta V_0^\parallel  (1,\ldots,N)]/Z_{\parallel},
\end{align}
which then yields
\begin{align}\label{eq:A4}
&\frac {N!}{(N-m)!}\frac {1}{Z_\parallel} \int \prod \limits _{j=m+1}^N dj \; \exp [-\beta V_0  (1,\ldots,N)]\nonumber\\
=&\rho^{(m)} _\parallel (1,\ldots, m;N) +\sum \limits _{1 \leq i < j \leq m} v_1(i,j) \rho_\parallel ^{(m)} (1,\ldots,m;N)\nonumber\\
& +\sum \limits _{i=1} ^m \int d(m+1) v_1 (i,m+1) \rho_\parallel ^{(m+1)}(1,\ldots,m+1;N)\nonumber\\
& + \frac{1}{2} \int d (m+1) \int d(m+2) v_1 (m+1, m+2)\nonumber\\
&\times \rho_\parallel ^{(m+2)} (1,\ldots, m+2;N)+\mathcal{O} (L^4).
\end{align}
Here, the variable $N$ of $\rho_\parallel ^{(m)} (1,\ldots,m;N)$ shall indicate, that the support of the $m$-particle densities is restricted only to the finite system of area $A$ with finite particle number $N$.
The partition function $Z(L)$ follows similarly, but does not require the decomposition of Eq.~\eqref{eq:A1}. Taking $\int d^\perp j=  \int dz_j \exp [-\beta \mathcal{U}(z_j;L)]/z_{\perp}(L)=1$ into account one infers with Eq.~\eqref{eq:seriespot} and $z_\perp ^N (L) = Z_\perp (L)$:
\begin{align}\label{eq:A5}
Z(L)&\nonumber \\
=& Z_{\perp}(L)\int \prod \limits _{j=1}^N dj \exp [- \beta V_0(1,\ldots,N)] \nonumber \\
=&Z_{\perp}(L) \int  \prod \limits ^N _{j=1} dj \exp[- \beta V_0^\parallel (1,\ldots, N)] \nonumber\\
&\times \left[1+\frac{1}{2}\frac{N!}{(N-2)!} v_{1}(1,2)+\mathcal{O}(L^4) \right]\nonumber \\
=& Z_\perp (L) Z_\parallel \negthinspace \left[1+\frac{1}{2} \int d 1 \int d2 v_1(1,2) \rho_\parallel ^{(2)} (1,2;N) +\negthinspace \mathcal{O} (L^4)\negthinspace\right].
\end{align}
Then the normalization factor (Eq.~\eqref{eq:normfactor}) becomes
\begin{equation}\label{eq:A6}
\mathcal{N}(L) = 1 - \frac{1}{2}  \int d1 \int d2 v_1(1,2) \rho_\parallel ^{(2)} (1,2;N) + \mathcal{O} (L^4).
\end{equation}
Substituting Eqs.~\eqref{eq:A4} and~\eqref{eq:A6} into Eq.~\eqref{eq:mpdensity-b} and renaming the integration variables in Eq.~\eqref{eq:A6} we arrive at
\begin{align}\label{eq:A7}
 \tilde{\rho}^{(m)}& (1,\ldots, m;L) \nonumber\\
=&\left[1+\sum \limits _{1 \leq i < j \leq m} v_1(i,j)\right]\rho_\parallel ^{(m)} (1,\ldots,m;N)\nonumber\\
&+ \sum \limits _{i=1} ^m \int d (m+1) v_1 (i,m+1) \rho_\parallel ^{(m+1)} (1,\ldots,m+1;N)\nonumber \\
&+ \frac 1 2 \int d (m+1) \int d (m+2)  v_1(m+1,m+2)\nonumber \\
& \times\Big[\rho_\parallel ^{(m+2)} (1,\ldots,m+2;N)\nonumber\\
&- \rho_\parallel ^{(m)} (1,\ldots,m;N) \rho_\parallel ^{(2)}(m+1,m+2;N)\Big]\nonumber \\
&+ \mathcal{O} (L^4).
\end{align}
Comparison with Eq.~\eqref{eq:seriesmpdensity} leads to
\begin{equation}\label{eq:A8}
\tilde{\rho} _0^{(m)} (1,\ldots,m;N;L) = \rho_\parallel ^{(m)} (1,\ldots,m;N),
\end{equation}
and
\begin{align}\label{eq:A9}
\tilde{\rho}_1^{(m)} &(1,\ldots,m;N;L)\nonumber \\
=& \sum \limits _{1 \leq i < j \leq m} v_1(i,j) \rho_\parallel ^{(m)}(1,\ldots,m;N) \nonumber \\
&+\sum \limits _{i=1} ^m \int d (m+1) v_1 (i,m+1) \rho_\parallel^{(m+1)} (1,\ldots,m+1;N) \nonumber \\
&+\frac 1 2 \int \diff (m+1) \int \diff (m+2) v_1(m+1,m+2)\nonumber \\
&\times[\rho_\parallel ^{(m+2)}(1,\ldots, m+2;N)\nonumber \\
&- \rho_\parallel ^{(m)} (1,\ldots,m;N)\rho_\parallel ^{(2)}(m+1,m+2;N)].
\end{align}
The first-order correction for $m=2$, Eq.~\eqref{eq:A9}, has the same structure as the correction obtained for a bulk fluid perturbed by a pair potential, 
see Ref.~\cite{Lowry:1964}. It has been pointed out in Ref.~\cite{Smith:1971} that this result is formal, i.e. valid for the finite system, only. 
To obtain its TD-limit we use a relation connecting the $m$-particle density of the finite system $\rho^{(m)}_{\parallel}(1,\ldots,m;N)$ to the corresponding quantity $\rho^{(m)}_{\parallel}(1,\ldots,m;\infty)\equiv\rho^{(m)}_{\parallel}(1,\ldots,m)$ of the infinite system~\cite{Lebowitz:1961}
\begin{align}\label{eq:A10}
\rho^{(m)}_{\parallel}&(1,\ldots,m;N)\nonumber \\
=& \rho^{(m)}_{\parallel}(1,\ldots,m) - \frac{1}{N}\left(\frac{n_0\kappa_T}{2\beta}\right) n_0^2\frac {\partial^2} {\partial n_0^2}\Big|_T \rho_\parallel ^{(m)} (1,\ldots,m)\nonumber \\
&+o(N^{-1}).
\end{align}
Here, $\kappa_T=\left[n_{0}\partial \Sigma / \partial n_{0}\Big|_{T} \right]^{-1}$ refers to the 2D isothermal compressibility of the reference fluid with surface tension given by $\Sigma=n_{0}k_{B}T+ k_{B}T\partial \ln (Z_{\parallel})/ \partial A \Big|_{T,N}$.

The second term in Eq.~\eqref{eq:A10}  is of order $1/N$, however it gives rise to an intensive contribution via the double integral of Eq.~\eqref{eq:A9}. Its generic contribution stems from  such $((m+1)^{\parallel},(m+2)^{\parallel})$ with a distance to the fixed positions  $(1^{\parallel},\ldots,m^{\parallel})$, which is much larger than the correlation length. In this case one can use in Eq.~\eqref{eq:A10} the factorization property
\begin{align}\label{eq:A12}
{\rho}_{\parallel}^{(m+2)} (1,\ldots,m+2) \rightarrow {\rho}_{\parallel}^{(m)} (1,\ldots,m) {\rho}_{\parallel}^{(2)} (m+1,m+2),
\end{align}
see Ref. ~\cite{Lebowitz:1961}.
As a final result we obtain
\begin{align}\label{eq:A13}
 \tilde{\rho }_1^{(m)}& (1,\ldots,m;L) \nonumber\\
 =& \sum \limits _{1 \leq i < j \leq m} v_1(i,j) \rho_\parallel ^{(m)}(1,\ldots,m)\nonumber\\
& + \sum \limits _{i=1} ^m \int d (m+1) v_1 (i,m+1) \rho_\parallel^{(m+1)} (1,\ldots,m+1) \nonumber \\
& + \frac 1 2 \int \diff (m+1) \int \diff (m+2) v_1(m+1,m+2)\nonumber \\
&\times \negthinspace\negthinspace \left[\rho_\parallel ^{(m+2)}\negthinspace(1,\ldots, m+2)\negthinspace-\negthinspace \rho_\parallel ^{(m)} \negthinspace(1,\ldots,m)\rho_\parallel ^{(2)}\negthinspace(m+1,m+2)\negthinspace\right] \nonumber\\
&-\left(\frac{n_0^2 \kappa_T}{2\beta }\right) \left(\frac {\partial}{\partial n_0} \Big|_T \rho^{(m)}_\parallel (1,\ldots,m)\right) \int \diff (m+2)\nonumber \\
&\times v_1(m+1,m+2)\frac {\partial}{\partial n_0} \Big|_T \rho^{(2)}_\parallel (m+1,m+2).
\end{align}
For the last term we have used that the integrand of the double integral, Eq.~\eqref{eq:A10}, depends on $|(m+2)^{\parallel}-(m+1)^{\parallel}|$ by translational invariance yielding an intensive term proportional to $A/N=1/n_0$.

\bigskip

\section{Calculation of the reduced $1$-particle density $\tilde{\rho}_{1}^{(1)}(\vec{r}_1 z_1;L)$}\label{Sec:bdensity}

We employ the shorthand notation as introduced in Appendix \ref{Sec:acum} to evaluate  Eq.~\eqref{eq:correctmpdensity-1} in the case of $m=1$
\begin{align}\label{eq:B1}
\tilde{\rho}^{(1)}_1&(1;L) \nonumber \\
=& \langle v_1^\perp (1,2)\rangle_\perp \int d^\parallel 2 v_1^\parallel (1,2) \rho_\parallel ^{(2)}(1,2)  \nonumber \\
&+ \frac 1 2 \langle v_1^\perp(2,3)\rangle _\perp \int d^\parallel 2 \int d^\parallel 3 v_1^\parallel (2,3)\nonumber \\
&\times\left[\rho_\parallel ^{(3)} (1,2,3) - \rho^{(1)}_\parallel (1) \rho^{(2)}_\parallel (2,3)\right]\nonumber \\
&-\langle v_1^\perp (2,3) \rangle _\perp \left(\frac {n_0^2 \kappa_T}{2 \beta}\right) \int d^\parallel 3 v_1^\parallel (2,3) \frac {\partial }{\partial n_0} \Big|_T \rho_\parallel ^{(2)}(2,3),
\end{align}
where $\frac {\partial}{\partial n_0} \Big|_T \rho_\parallel ^{(1)} (1) \equiv 1$ has been used, since $\rho_\parallel ^{(1)} (1) \equiv n_0$. Substituting $\frac {\partial}{\partial n_0} \Big|_T \rho_\parallel ^{(2)} (2,3)$ from Eq.~\eqref{eq:densityderiv} into Eq.~\eqref{eq:B1} yields for the last term of Eq.~\eqref{eq:B1}
\begin{align}\label{eq:B2}
&- \langle v_1^\perp (2,3) \rangle _\perp \Big\{ \int d^\parallel 3 v^\parallel _1 (2,3) \rho_\parallel ^{(2)} (2,3) \nonumber\\
&+ \frac 1 2 \int d^\parallel 3 v_1^\parallel (2,3) \int d^\parallel 4  [\rho^{(3)}_\parallel (2,3,4) -  \rho^{(2)}_\parallel (2,3)\rho^{(1)}_\parallel (4) ] \Big\}.
\end{align}
Consequently we obtain after renaming dummy variables:
\begin{align}\label{eq:B3}
 \tilde{\rho}^{(1)}_1 &(1;L)\nonumber \\
=& [\langle v_1^\perp (1,2)\rangle _\perp - \langle v_1^\perp (2,3)\rangle _\perp ] \int d^\parallel 2 v^\parallel _1 (1,2) \rho_\parallel^{(2)} (1,2) \nonumber  \\
& + \frac 1 2 \langle v_1^\perp (2,3)\rangle _\perp \Big\{\int d^\parallel 2 d ^\parallel 3 v_1^\parallel (2,3)\nonumber\\
&\times \left[\rho^{(3)}_\parallel (1,2,3) - \rho ^{(1)}_\parallel (1) \rho^{(2)}_\parallel (2,3)\right] \nonumber \\
&-\int d^\parallel 3 d ^\parallel 4 v_1^\parallel (2,3) \left[\rho^{(3)}_\parallel (2,3,4) - \rho^{(1)}_\parallel (4) \rho ^{(2)}_\parallel (2,3)\right]\Big\}.
\end{align}

By translational invariance, the first term of the curly bracket in Eq.~\eqref{eq:B3} does not depend on $1 ^\parallel \cong  \vec{r}_1$ and the second term is independent of $2 ^\parallel \cong \vec{r}_2$. After renaming the label $4 \to 1$ and employing the permutation invariance of $\rho_\parallel ^{(3)}(1,2,3)$ one infers
\begin{align}\label{eq:B4}
 \lim_{A\to \infty}& \Big[\frac{1}{A}\int d^\parallel 1 \int d ^\parallel 2 \int d^\parallel 3 v_1^\parallel (2,3)\nonumber \\
&\times(\rho_\parallel ^{(3)} (1,2,3) - \rho^{(1)}_\parallel (1) \rho^{(2)}_\parallel (2,3))  \nonumber \\
& -\frac{1}{A}\int d^\parallel 2 \int d^\parallel 3 \int d^\parallel 4 v_1^\parallel (2,3)(\rho^{(3)}_\parallel (2,3,4)\nonumber\\ 
&- \rho^{(1)}_\parallel (4) \rho^{(2)}_\parallel (2,3))\Big]= 0.
\end{align}
Finally, with Eq.~\eqref{eq:B4} it follows
\begin{align}\label{eq:B5}
\tilde{\rho}_1^{(1)} (1;L) =& \left[\langle v_1^\perp (1,2)\rangle _\perp - \langle v_1^\perp (2,3)\rangle _\perp \right] \nonumber \\
&\times\int d^\parallel 2 v_1^\parallel (1,2) \rho_\parallel ^{(2)}(1,2),
\end{align}
which is independent of $1 ^\parallel \cong \vec{r}_1$.


\section{Canonical averages of the cluster function of hard spheres}\label{Sec:ccluster}

In the case of hard-sphere interaction the first-order correction $\tilde\rho^{(m)}_{1}$ contains integrals of the subsequent form
\begin{align}\label{eq:C1}
I_1^{(m)}&(z_1,\vec{r}_1,\ldots,\vec{r}_m;L) \nonumber\\
=&\int dz_{m+1}\rho^{(1)}_{\perp}(z_{m+1})\int d^{2}r_{m+1}f(\vec{r}_1z_1,\vec{r}_{m+1}z_{m+1})\nonumber \\
&\times \rho_\parallel^{(m+1)}(\vec{r}_1,\ldots,\vec{r}_{m+1};\sigma_{L}),
\end{align}
where $\rho_\parallel^{(m)}(\vec{r}_1,\ldots,\vec{r}_{m};\sigma_{L})$ refers to the 2D $m$-particle density of hard disks with reduced diameter $\sigma_{L}$.
Inserting for $f(\vec{r}_1z_1,\vec{r}_{m+1}z_{m+1})$ the explicit representation of the cluster-function from Eq.~\eqref{eq:clusterf} and using $\vec{r}_{m+1}=\vec{r}_{1}+\vec{r}_{1,m+1}$ one obtains
\begin{align}\label{eq:C2}
&I_1^{(m)}(z_1,\vec{r}_1,\ldots,\vec{r}_m;L)\nonumber\\
=&\int dz_{m+1}\rho^{(1)}_{\perp}(z_{m+1})\int_{0}^{2\pi} d\varphi_{m+1}  \int_{0}^{\infty} dr_{1,m+1} r_{1,m+1}\nonumber\\
&\times\big[\Theta(r_{1,m+1}^2+(z_{1}-z_{m+1})^2-\sigma^2)-\Theta (r_{1,m+1}^2-\sigma_{L}^2) \big] \nonumber\\
&\times\rho_\parallel^{(m+1)}(\vec{r}_1,\ldots,\vec{r}_1+ r_{1,m+1}\vec{e}(\varphi_{1,m+1});\sigma_{L}) \nonumber\\
=&-\int dz_{m+1}\rho^{(1)}_{\perp}(z_{m+1})\int_{0}^{2\pi} d\varphi_{m+1}  \negthinspace\negthinspace \int_{\sigma_{L}}^{\sqrt{\sigma^2-(z_{1}-z_{m+1})^2}}\negthinspace  \negthinspace\negthinspace\negthinspace\negthinspace \negthinspace\negthinspace dr_{1,m+1}\nonumber\\
& \times r_{1,m+1} \rho_\parallel^{(m+1)}(\vec{r}_1,\ldots,\vec{r}_1+ r_{1,m+1}\vec{e}(\varphi_{1,m+1});\sigma_{L}),
\end{align}
where the unit vector $\vec{e}(\varphi_{1,m+1})=\vec{r}_{1,m+1}/r_{1,m+1}$ has been introduced. The $r_{1,m+1}$-integration interval  converges to zero for $L\to 0$. Therefore one can expand the integral around its lower bound
\begin{align}\label{eq:C3}
I_1^{(m)}&(z_1,\vec{r}_1,\ldots,\vec{r}_m;L) \nonumber\\
=&-\frac{1}{2}\int dz_{m+1}\rho^{(1)}_{\perp}(z_{m+1})\left[\sigma^2-(z_{1}-z_{m+1})^2 -\sigma_{L}^2 \right]\nonumber\\
&\times \negthinspace\negthinspace\int_{0}^{2\pi} \negthinspace\negthinspace d\varphi_{m+1} \rho_\parallel^{(m+1)}(\vec{r}_1,\ldots,\vec{r}_1 \negthinspace\negthinspace+ \negthinspace\negthinspace \sigma_{L}\vec{e}(\varphi_{m+1});\sigma_{L})+\mathcal{O}(L^4),
\end{align}
where $\varphi_{1,m+1}=\varphi_{m+1}$ without restricting generality.
With $\sigma_{L}=\sqrt{\sigma^2-L^2}$  one finally obtains
\begin{align}\label{eq:C4}
I_1^{(m)}&(z_1,\vec{r}_1,\ldots,\vec{r}_m;L)\nonumber\\
=& \pi \left[z_{1}^2-2z_{1}\langle z_{m+1} \rangle_{\perp}+  \langle z_{m+1}^2 \rangle_{\perp}-L^2 \right]\nonumber\\
&\times\langle\rho_\parallel^{(m+1)}(\vec{r}_1,\ldots,\vec{r}_1+ \sigma_{L}\vec{e}(\varphi_{m+1});\sigma_{L}) \rangle_{\varphi} \nonumber\\
& + \mathcal{O}(L^4),
\end{align}
where angular averages are indicated by
\begin{eqnarray}\label{eq:C5}
\langle(\cdots)\rangle_{\varphi}= \left(\frac{1}{2\pi}\right)\int_{0}^{2\pi}  \diff \varphi (\cdots).
\end{eqnarray}
The square bracket in Eq.~\eqref{eq:C4} corresponds to the first three terms occurring in Eq.~\eqref{eq:climit}, which arises from $\langle v_1^{\perp}(z_1,z_2)\rangle_{\perp}$ except for the term $L^2$ which stems from the diameter $\sigma_{L}$ of the reference fluid.



%

\end{document}